# On-demand longitudinal transformations of light beams via an amplitude-phase-decoupled Pancharatnam–Berry phase element

Yu Zou[†], Yanke Li[†], Jialong Yan, Hui Liu, Yuqi Zhang, Dandan Wen, Peng Li, Bingyan Wei,[*] Jianlin Zhao, and Sheng Liu[**]

*Key Laboratory of Light Field Manipulation and Information Acquisition, Ministry of Industry and Information Technology, and Shaanxi Key Laboratory of Optical Information Technology, School of Physical Science and Technology, Northwestern Polytechnical University, Xi'an 710129, China*

* Corresponding author: wbyxz@nwpu.edu.cn

** Corresponding author: shengliu@nwpu.edu.cn

[†]These authors have contributed equally to this work and share the first authorship

**Abstract:** Overcoming the inherent phase-only modulation limitation of conventional Pancharatnam-Berry (PB) elements, we propose a checkerboard-encoded PB element that enables simultaneous amplitude and phase modulation through PB phase engineering. This single-element platform directly generates polarization-rotating beams with continuously varying polarization angles along the propagation axis, eliminating traditional dynamic phase requirements. By strategically combining the proposed element with a liquid crystal $q$-plate and waveplates, we demonstrate on-demand longitudinal transformations of polarization states and high-order Poincaré sphere beams. Significantly, we achieve, for the first time to our knowledge, the propagation-transformed optical skyrmions, whose topological textures evolve with propagation distance. This amplitude-phase decoupling mechanism and the integration solution for various longitudinal transformations open new avenues in structured light manipulation, particularly benefiting quantum optics and topological photonics applications requiring longitudinal mode control.

## 1. Introduction

In optics, the Pancharatnam-Berry (PB) phase describes the geometric phase acquired during polarization transformations [1, 2]. This phenomenon provides an additional degree of freedom for manipulating polarization states of light fields and designing advanced optical components. In contrast to conventional dynamic phase optical elements [3, 4], which modulate phase through optical path differences achieved by altering device geometry or medium refractive index, PB phase elements (PBPEs) control phase by adjusting optical axis distributions to regulate polarization states. Such elements reverse the handedness of the circularly polarized beams while attaching opposite PB phases corresponding to their polarization states [5, 6]. Typical PBPE implementations include subwavelength gratings [7, 8], liquid crystals (LC) [9-12], metasurfaces [13-16], etc. These components find extensive applications in polarization modulation, including the generation [17-21] and enhancement of spin Hall effects [22-24], as well as producing vector beams with various polarization distributions [25-28]. All these polarization modulation effects are achieved by PBPE-mediated control of the transverse polarization structures of optical beams.

Recent advances in optical field manipulation have propelled significant interest in structured beams with propagation-dependent characteristics, particularly regarding their potential for multidimensional information encoding. The longitudinal varying features of optical beams, including longitudinally varying intensities [29], trajectories [30], orbital angular momentum (OAM) [31,32], polarization states [33, 34], transverse mode [35], and even synthesized combinations thereof [36], have been systematically explored through various photonic platforms. Although spatial light modulators have become the predominant solution for longitudinal modulation in polarization state [37-41], their dependence on cascaded components increases system complexity. LC PBPEs [10,11,42] offer a more compact alternative. However, conventional PBPEs remain constrained by their inherent limitation of supporting only the spin-dependent PB phase modulations while lacking amplitude control capabilities, necessitating auxiliary components like Bessel beam generators [43]. While metasurfaces can overcome this limitation through additional amplitude and dynamical phase modulations via structural design [33, 44], they face challenges in large-area fabrication costs and optical efficiency optimization compared to LC-based systems. This underscores the critical need for developing single-LC-PBPE solutions capable of decoupling the modulation of amplitude and PB phase.

Following the diversified development in longitudinal manipulations of optical fields in recent years, applications have expanded into stereo polarization holography [45], tunable refractometry [46], and nonreciprocal optical devices [41]. However, several key challenges remain unresolved. For example, the effective longitudinal modulation of topologically structured optical modes (e.g. optical skyrmions [47]) remains unrealized due to their inherent topological robustness; overcoming this limitation would introduce a novel topological degree of freedom to optical field manipulation. Additionally, existing longitudinal modulation mechanisms lack a unified theoretical framework and technical platform. Consequently, establishing compatible and extensible longitudinal modulation methods represents a critical challenge in optical field manipulation research.

In this paper, we propose a LC PBPE based on the checkerboard encoding design to simultaneously modulate the amplitude and the PB phase. This LC element is fabricated by the SD1-based photoalignment technology and the digital micro-mirror device (DMD) micro-lithography system. The polarization-rotating beams (PRB), i.e., the beams whose polarization varies along the equator of the Poincaré sphere, can be directly generated with a single checkerboard-type LC PBPE. When used in conjunction with a quarter-wave plate (QWP) and a $q$-plate, the system can produce scalar and vector beams that vary along different paths on the Poincaré sphere and higher-order Poincaré spheres. Additionally, by adjusting the voltage of the $q$-plate, it is possible to generate Skyrmions textures that vary along the Skyrmions ring with increasing propagation distance, further expanding the dimensions of structured light control. This compact integration solution would not only improve experimental system flexibility but could also facilitate the development of a universal theoretical framework for longitudinal modulation.

## 2. Theoretical Principle

Ideally, a stable PRB propagating along the $z$-axis can be expressed by superposing two circularly polarized Bessel beams with different radial wave vectors [43]

$$\mathbf{E}_{\mathrm{PRB}} = J_0(k_{r+}r)\mathbf{E}_{\mathrm{L}} + J_0(k_{r-}r)\mathrm{e}^{-\mathrm{i}2k_r\delta_k z/k_0}\mathbf{E}_{\mathrm{R}}, \qquad (1)$$

where $J_0(\cdot)$ is the zero-order Bessel function of the first kind; $k_{r\pm}=k_r\pm\delta_k$, $k_r$ is the radial component of wave vector $k_0=2\pi/\lambda$, $\delta_k$ denotes the radial variation of the wave vector; $\mathbf{E}_{\mathrm{L/R}}=[1, \pm\mathrm{i}]^\mathrm{T}/\sqrt{2}$ denote the left- and right-handed circular polarizations, respectively; $\Delta\varphi_z \approx 2k_r\delta_k z/k_0$ is the difference of the Gouy phase shift between left- and right-handed components. The on-axis polarization can be expressed by

$\mathbf{E}_{\text{PRB}}(r=0)=[\cos(\Delta\varphi_z/2),\sin(\Delta\varphi_z/2)]^{\text{T}}$. Thus, the rate of change of the beam polarization can be expressed as $\alpha = k_r\delta_k/k_0$, and the corresponding period of change is $z_T=2\pi/\alpha$.

According to the equivalence between the Bessel beam $J_0(k_{r\pm}r)$ and the conical wave $\exp(-ik_{r\pm}r)$ to some extent in an experiment, we can extract a common term in Eq. (1), and get

$$\mathbf{E}_{\text{PRB}(z=0)} = J_0(k_r r)\left[e^{i\delta_k r}\mathbf{E}_{\text{L}} + e^{-i\delta_k r}\mathbf{E}_{\text{R}}\right]. \qquad (2)$$

It is as if a linear polarized Bessel beam passes through a PBPE. For the conventional PBPE (assuming its optical axis distribution is $\delta_k r/2$), it can transform the left- and right-handed components into right- and left-handed ones and attach different phases $\pm\delta_k r$, respectively. Therefore, to realize the PRB, another element (e.g., conical-phase elements) is needed to transform the amplitude of the beam into a Bessel profile, which cannot be realized by a single PBPE so far.

Here, the checkerboard encoding technology is adopted in the design of the PBPE to realize the amplitude and PB phase modulation simultaneously. Figure 1(b) shows the principle of the checkerboard phase structure design of the PBPE. Phases $\varphi_1$ and $\varphi_2$ are loaded into two complementary checkerboard grids, where the colored squares denote the loaded phases and the black squares are null. These two-phase structures combine to form a checkerboard pattern with alternating phases $\varphi_1$ and $\varphi_2$. Theoretically, the light field modulated by the checkerboard-based phase is the superposition of different diffraction orders, of which the diffraction angles and intensities are related to the checkerboard cell width $d_l$. The zero-order diffraction field can be expressed as $\exp(i\varphi_1)+\exp(i\varphi_2)=2\cos[(\varphi_1-\varphi_2)/2]\exp[i(\varphi_1+\varphi_2)/2]$ [49]. When a horizontally polarized ($[1, 0]^{\text{T}}$) plane wave passes through this PBPE [see Fig. 1(a)], the right- and left-handed polarization components of the light field become $\exp(\pm i\varphi_1)+\exp(\pm i\varphi_2)$, respectively. Then, the output field is expressed as

$$\begin{aligned}\mathbf{E}_{\text{out}(z=0)} &= \frac{1}{\sqrt{2}}\left[\left(e^{i\varphi_1}+e^{i\varphi_2}\right)\mathbf{E}_{\text{L}}+\left(e^{-i\varphi_1}+e^{-i\varphi_2}\right)\mathbf{E}_{\text{R}}\right] \\ &= \sqrt{2}\cos\left(\frac{\varphi_1-\varphi_2}{2}\right)\cdot\left(e^{i\frac{\varphi_1+\varphi_2}{2}}\mathbf{E}_{\text{L}}+e^{-i\frac{\varphi_1+\varphi_2}{2}}\mathbf{E}_{\text{R}}\right)\end{aligned}. \qquad (3)$$

From Eq. (3), the checkerboard-encoded PBPE attaches the left- and right-handed components with opposite phases $\pm i(\varphi_1+\varphi_2)/2$ and modulates the amplitude into $\cos[(\varphi_1-\varphi_2)/2]$. To produce the beam specified in Eq. (2), we simply need to set $\cos[(\varphi_1-\varphi_2)/2]=J_0(k_r r)$ and $(\varphi_1+\varphi_2)/2=\delta_k r$. Yet, this model can be further simplified. According to the stationary phase approximation, the zero-order Bessel function can be written as $J_0(k_r r) \approx \sqrt{2/(\pi k_r r)} \cos(k_r r - \pi/4)$ for a large $r$. Actually, the cosine amplitude can be resolved into diverging and converging conical wave components $\exp[\pm i(k_r r - \pi/4)]$. The diverging part generates a Bessel beam, while the converging part does not affect axial intensity. In addition, the $r$-dependent term $\sqrt{2/(\pi k_r r)}$ should be removed, which would not affect the formation of the Bessel profile. Since the constant phase can be ignored, the amplitude of Eq. (2) can be replaced by a cosine function. Namely, $\mathbf{E}_{PRB(z=0)} = \cos(k_r r)[\exp(i\delta_k r)\mathbf{E}_L + \exp(-i\delta_k r)\mathbf{E}_R]$. Combining with Eq. (3) and ignoring the constant amplitude, the phases of the checkerboard-encoded PBPE can be solved as

$$\begin{cases} \varphi_1 = k_{r+} r \\ \varphi_2 = -k_{r-} r \end{cases}. \qquad (4)$$

According to the phase map (i.e., twice the optical axis direction) given in Eq. (4), we use the SD1-based optical alignment technique and DMD-based (resolution: 1920×1080; pixel size: 10.8 mm×10.8 mm) microlithography system to fabricate a checkerboard-encoded PBPE in LC. A 5× objective lens is utilized in the microlithography system to achieve a written pixel size of 2.16 mm. Considering the cell size of the checkerboard structure will affect the experimental results, we fabricate several samples with different cell sizes. Figure 1(c) shows the photograph of the six fabricated LC samples, of which the unit cells occupy 1, 2, 4, and 6 pixels, respectively. While the phase maps of the samples appear similar in photographs, their performance for producing PRB is much different, depending on the imaging accuracy and the quality of the LC plate. Decreasing the checkerboard cell size might reduce the disturbing of the higher-order diffraction. However, the smaller the cell size is, the easier it is for the LC structure in a unit cell to be affected by the adjacent cells. Thus, we choose the best one whose unit cell is 4-pixel wide (i.e. 8.64 mm) as shown in the bottom right of Fig. 1(c). The relevant parameters are $k_r = 2.5 \times 10^5$ m$^{-1}$ and $\delta_k = 30\pi$ mm$^{-1}$, theoretically producing a PRB with polarization rotation rate $a = 136.44°$/mm and rotation period $z_T = 2.64$ mm. Fig. 1(d) presents a polarized optical microscope micrograph of the checkerboard-encoded LC PBPE (about 2.3 mm in square). Fig. 1(e) offers a partial zoom view showcasing the visible checkerboard structure on the LC plate.

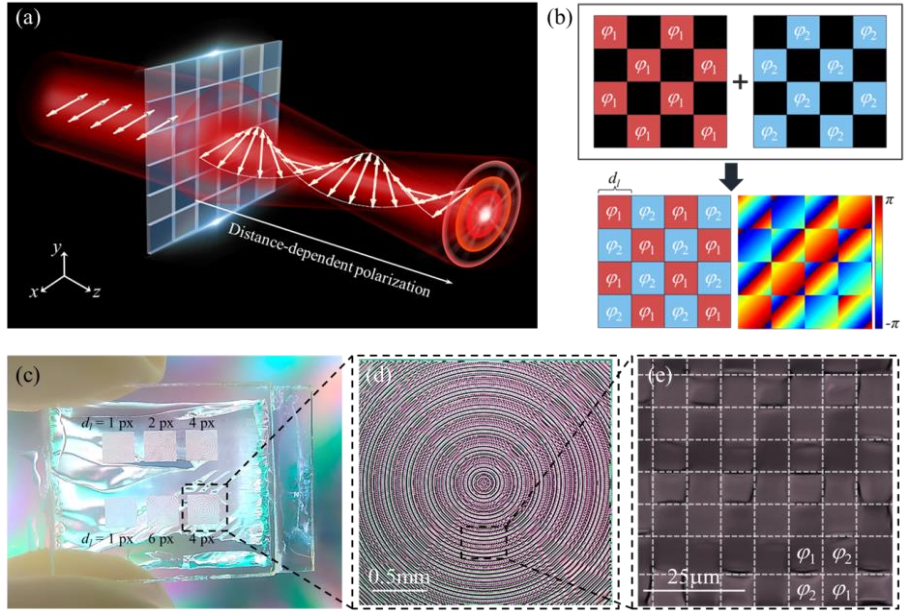

**Figure 1.** (a) Schematic of the generation of polarization-rotating beams (PRB) by a single checkerboard-encoded Pancharatnam-Berry phase element (PBPE). (b) Phase structure design of the PBPE. (c) Photograph of the fabricated samples taken with two crossed polarizers. (d, e) Polarization micrograph and its zoom view of the checkerboard-encoded liquid-crystal (LC) PBPE in the bottom right of (c).

## 3. Results and Discussion

**A. Generating the polarization-rotating beam**

To verify the above theory, we employ the experimental setup as shown in Fig. 2(a). The linearly polarized He-Ne laser beam (632.8 nm) from the polarizer P1 passes through the checkerboard-encoded LC PBPE and is directly transformed into a PRB. To improve the beam quality, a 4f system composed of lenses L1 and L2 is employed to filter out the higher-order diffraction and only the zero-order field passes through (see the white dashed circle in the inset). The conversion efficiency of the component is 43% (more details can be found in Section S1, Supporting Information). The intensity distribution of the beam is captured by a CCD camera moved step by step. A linear polarizer P2 and a QWP are placed before the CCD to acquire four intensity measurements at specific angular orientations, which are then used to calculate the Stokes parameters. The transverse intensity distributions at different propagation distances within one polarization rotating period are shown in Fig. 2 (b). The top and bottom rows represent the analyzed patterns by polarizer P2 when the side lobes and central main lobe are extinct, respectively, with the polarizer directions marked by the white dashed lines. The polarization rotates 360° from $z_1$ to $z_9$. To analyze polarization rotation quantitatively, the CCD camera moves along the z-direction in 0.1 mm steps within 13.2 mm. In Fig. 2(c), the propagation of horizontally (top) and vertically (bottom) polarized components reveals periodic alternating extinctions, basically consistent with the simulation results (see Fig. S2 in Supplementary Information), indicating polarization rotation. In Figure 2(d), the variation trends of the Stokes parameters over one period are illustrated. Parameters $S_1$ and $S_2$ oscillate between -1 and 1, while $S_3$ remains zero. These results demonstrate that the polarization state undergoes periodic variation along the equatorial plane of the Poincaré sphere.

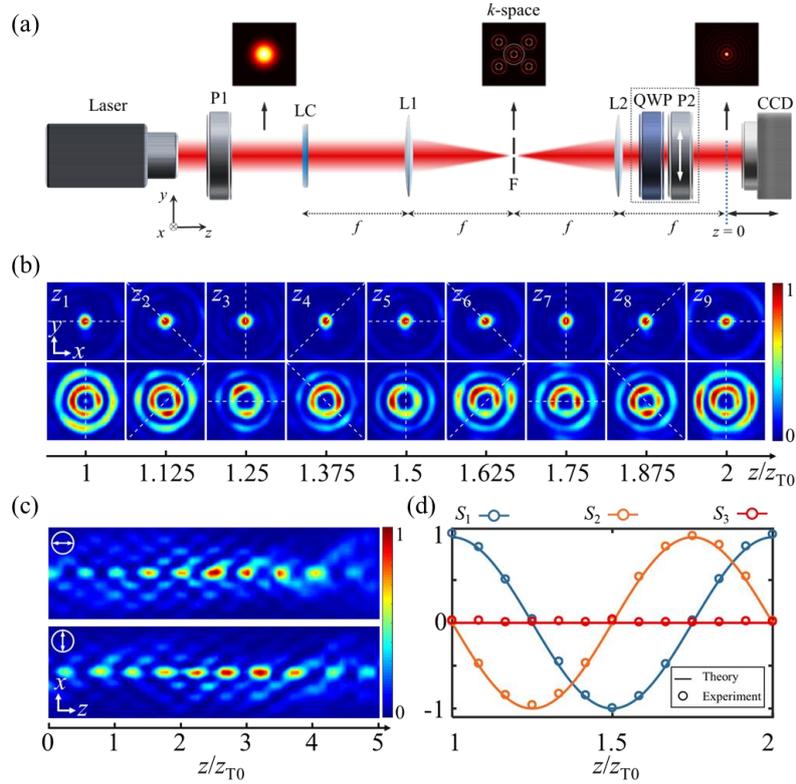

**Figure 2.** (a) Experimental setup. Laser, He-Ne laser at 632.8 nm; P1 and P2, polarizer; LC, checkerboard-encoded PBPE; L1 and L2, lens; F, aperture filter; QWP, a quarter-wave plate; Insets from left to right: input Gaussian beam, spatial spectrum distribution, and output intensity distribution. (b) Intensity distributions of the PRB analyzed by polarizers along the white dashed lines, at different distances within one period. (c) Side view of the propagation process of the horizontally and vertically polarized components. (d) The variation trends of the Stokes parameters over one period.

Figure 3(a) shows the relationship curve of the polarization rotation angle $\theta$ and the distance $z$ within 13.2 mm, where the blue solid line and the orange circles represent the theoretical and experimental results, respectively. The inset gives the zoom view of the results measured in one rotation period. It can be seen that the polarization rotation angle $\theta$ linearly changes with the propagation distance $z$. The experimental and theoretical results match well. It should be noted that the averaged rotation rate of the beam (the slope of the curve) in the experiment is about 137.0°/mm, with a relative error of 0.4% compared to the theoretical value. A beam expander can also be added to the system to flexibly control the polarization rotation rate of the emitted beam (more details can be found in Section S3, Supporting Information). To demonstrate the stability of the rotation rate of the resulting beam, we calculate the relative error in the measured rotation rate versus the propagation distance, as shown in Fig. 3(b). The average error and the root mean square error of the measured data are 0.6°/mm (0.4%) and 1.6°/mm (1.2%), respectively, indicating that the fluctuation of the polarization rotation rate is very small. These errors might be caused by the imaging quality of the DMD system during the fabrication process of the LC sample, the quality of the input beam, or the information loss in spatial filtering.

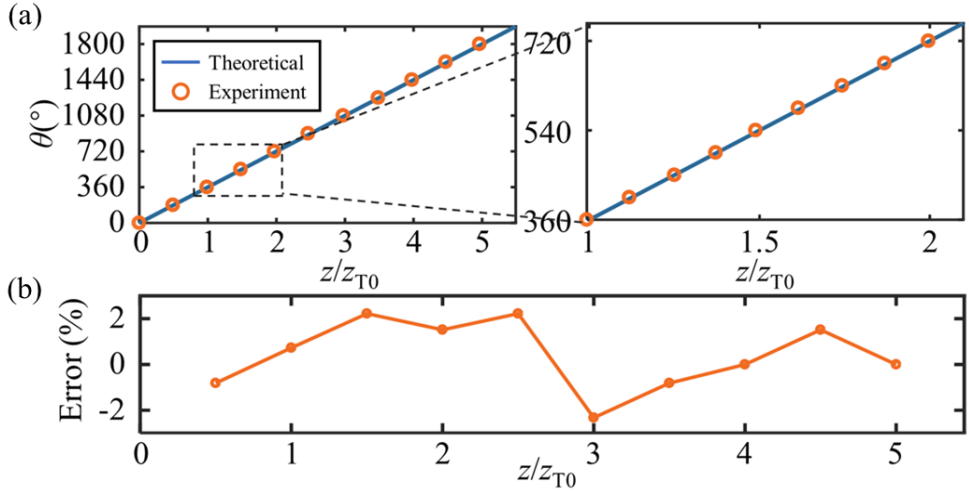

**Figure 3.** (a) Measured rotation angle $\theta$ vs. the propagation distance $z$, where the inset shows the zoom view within one period; (b) Relative error of the measured rotation rate vs. propagation distance.

### B. Longitudinal transformation of polarization through the PBPE

To achieve versatile longitudinal polarization modulation, we introduce a QWP as a polarization tailoring element into the baseline optical configuration. When an $x$-polarized polarized light sequentially traverses the QWP and the proposed LC element, the ellipticity $\Theta$ and azimuth angle $\Phi$ of on-axis polarization of the output beam are

$$\Theta = -\beta$$
$$\Phi = -\beta - \frac{\Delta\varphi_z}{2} \tag{5}$$

where $\beta$ is the angle between the fast axis of the QWP and the horizontal plane. The azimuthal angle $\Phi$ varies with z, indicating polarization rotation of the beam. The ellipticity is adjusted via $\beta$. By adjusting the relative position of QWP and LC element, the polarization evolution path can be further controlled. When a QWP is placed after the LC element, the ellipticity $\Theta$ and azimuth angle $\Phi$ of the output beam become:

$$\Theta = \frac{\Delta\varphi_z}{2} - \beta$$
$$\Phi = -\beta \quad (6)$$

As indicated by the formula, when $\beta$ is a constant, the ellipticity of the output light varies with the propagation distance z. On the Poincaré sphere, this evolution corresponds to movement along a meridian (more details can be found in Section S4 A, Supporting Information). In the experiment, with $\beta=-22.5°$, the change path of polarization state on the Poincaré sphere is shown by the red line in Figure 4 (a). Placing the QWP after the LC element results in a change of the polarization state along the meridian of the Poincaré sphere, as indicated by the blue line in Figure 4(a). Figures 4(b) and (c) respectively show the variation trends of the Stokes parameters as the beam polarization evolves along these tow paths during one transmission period. The lines in the figures represent theoretical results and the circles represent experimental results, which are in good agreement.

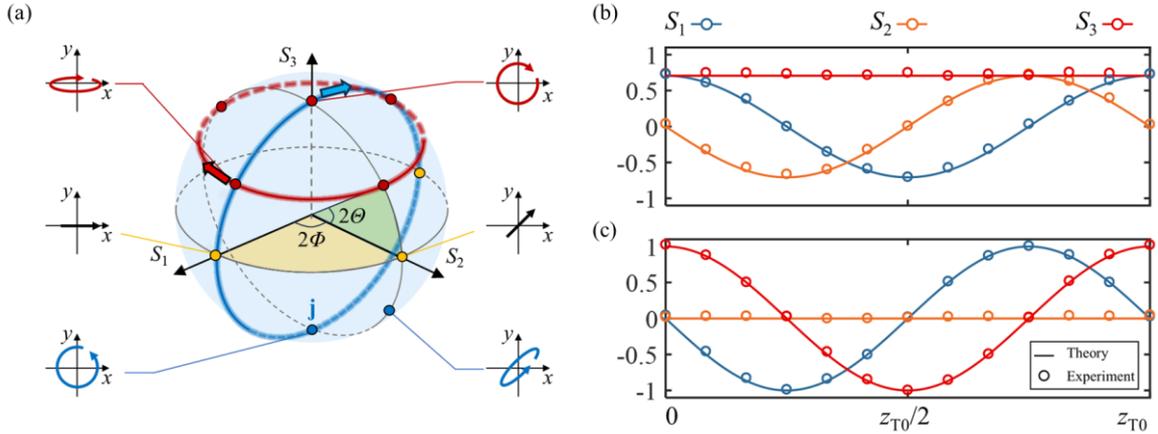

**Figure 4.** (a) The two transformation paths of polarization states (red, and blue lines) on the Poincaré sphere. (b) and (c) correspond to the variations in the Stokes parameters for these two paths.

## C. Longitudinal transformation of higher-order Poincaré sphere beams

The conventional Poincaré sphere effectively characterizes light fields with spatially homogeneous polarization distributions but exhibits inherent limitations when addressing vector beams featuring transversely inhomogeneous polarization structures. To overcome this constraint, the Higher-Order Poincaré Sphere (HOPS) [50-52] model has been proposed, which incorporates OAM as an additional dimension to geometrically represent complex spatially varying polarization states. To verify the modulation capability of the proposed LC element on vector beams, we construct a hybrid control system comprising a QWP, the proposed LC element and a $q$-plate. On the Poincaré sphere, the initial polarization state exhibits a clockwise evolution along both parallels and meridians. The effect of a waveplate is equivalent to a rotation by a specific angle around the Poincaré sphere axis aligned with its fast axis. The introduction of a $q$-plate imposes spin-orbit interaction, converting spin angular momentum into OAM. This process maps the initial uniform polarization state to a vector field on the HOPS, while simultaneously rotating the operational axis by $\pi$ around the $S_1$ axis, thereby reversing the polarization evolution trajectory. When the polarization state evolves along the latitude lines on the higher-order Poincaré sphere, the ellipticity and azimuth angle are characterized as follows:

$$\Theta = \beta$$
$$\Phi = \beta + \frac{\Delta\varphi_z}{2} \quad (7)$$

When the polarization state evolves along the meridians (longitude lines) on the higher-order Poincaré sphere, the ellipticity and azimuth angle are characterized as follows (more details can be found in Section S4 B, Supporting Information):

$$\Theta = \beta - \frac{\Delta\varphi_z}{2}$$
$$\Phi = \beta \quad (8)$$

In the experiment, we use a $q$-plate with $q=1/2$ to generate the first-order vector vortex beam. The longitudinal variation of polarization state along different paths on the HOPS is achieved using the arrangement and combination of the above components, as shown by the yellow, red, and blue lines in Figure 5a. Figures 5 (b) - (d) show the longitudinal evolution characteristics of beam polarization states under the three typical modulation modes. Figure 5b shows the variation of the polarization state of a beam on the equator of the HOPS, which undergoes a transition from an angle polarized beam to a radial polarized beam, and then to an angle polarized beam as the propagation distance increases. By rotating $\beta=22.5°$, Figure 5 (c) shows the polarization state of the beam changing along the high latitude path of the HOPS. In Figure 5 (d), the polarization state of the beam changes along the HOPS meridian.

Reordering the elements can also generate hybrid polarization vector beams with longitudinal variations (see Fig. S5 in Supplementary Information).

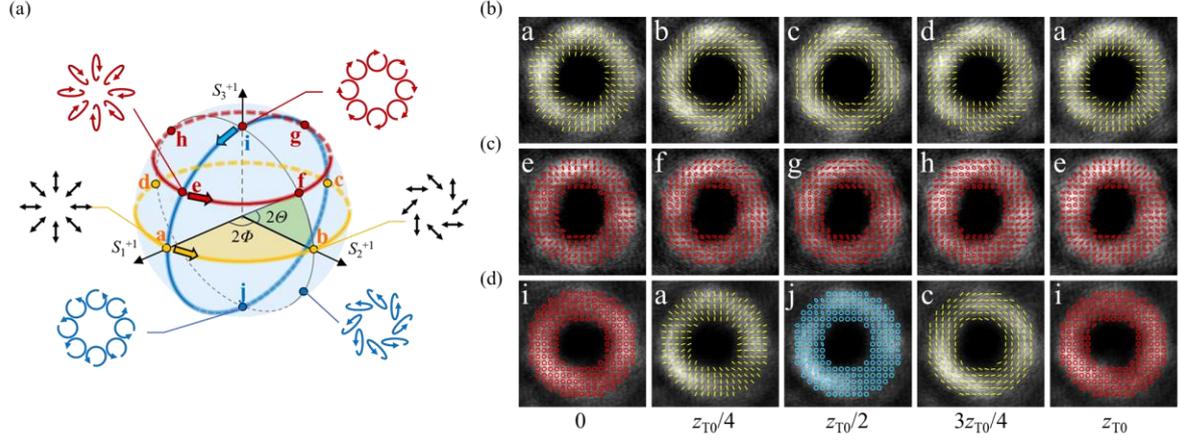

**Figure 5.** (a) Three characteristic evolution trajectories (yellow, red, blue lines) mapped on the higher-order Poincaré sphere (HOPS) surface. (b) Equatorial, (c) mid-latitude, and (d) longitudinal polarization transitions with propagation.

**D**. **Longitudinally varying optical skyrmions**

Skyrmions are a type of quasiparticle with distinct vector structures and topological protection characteristics, found in nuclear and condensed matter physics [53]. With the rapid development of topological optics, the existence of optical skyrmions has been confirmed by the scientific community [47], pioneering a new research direction for optical field control. They hold potential application value in fields such as micro-displacement sensing [54] and optical communication encryption [55].

The topological properties of skyrmion can be represented by the skyrmion number $s$, which is defined by [56]

$$s = \frac{1}{4\pi} \iint_A \mathbf{n} \cdot \left( \frac{\partial \mathbf{n}}{\partial x} \times \frac{\partial \mathbf{n}}{\partial y} \right) \mathrm{d}x\mathrm{d}y, \qquad (9)$$

in which $\mathbf{n}(x,y)$ is the vector field and $A$ is the area confining the skyrmion. Optical skyrmions in free space can be constructed through the Stokes parameters of structured vector beams. Then the vector field $\mathbf{n}(x,y)$ can be represented as $\mathbf{n}(x,y)= (S_1, S_2, S_3)$. The topological state of skyrmions is jointly regulated by multiple parameters such as the topological number, polarity, vorticity, and helicity angle. Their typical spatial textures can be divided into Néel type, Bloch type, and Anti-skyrmion type. Researchers have achieved the mutual transformation of topological textures in a plane by regulating the Stokes vector field and systematically characterized the texture transformation using the skyrmion torus model [56].

Conventional skyrmion interconversion requires manual hologram switching via spatial light modulators. Our element demonstrates the first realization of autonomous longitudinal transformation between optical skyrmion textures during propagation. By employing a LC $q$-plate operating at non-half-wave voltages, we generate full Poincaré beam profiles [57], which can be used to construct optical skyrmions (more details can be found in Section S6, Supporting Information). For example, Néel and Bloch skyrmions correspond to full Poincaré beam beams with lemon-type C-point, while antiskyrmions correspond to star-type C-point. Traditional skyrmions traversing our LC element undergo modulation, transforming into longitudinally self-switching skyrmions. This can be expressed as

$$\mathbf{E}_{out} = \sin\left(\frac{\delta_\phi}{2}\right) J_{\mp 1}(k_{r_+} r) \mathbf{E}_L + \cos\left(\frac{\delta_\phi}{2}\right) J_0(k_{r_-} r) e^{-i\Delta\varphi_z} \mathbf{E}_R \qquad (10)$$

where $J_0(\cdot)$ and $J_{\pm 1}(\cdot)$ respectively represent the zero-order and first-order Bessel modes, and $\delta_\phi$ is the optical phase retardation. The experimental results are shown in Figure 6. Figure 6(a) quantitatively characterizes the evolution trajectory of skyrmion textures on the toroidal surface during propagation, with the left and right insets illustrating theoretical diagrams of two distinct transformation pathways. Figures 6(b)-(c) demonstrate the transverse polarization distributions (upper panels) and their corresponding skyrmion textures (lower panels) at distinct propagation distances. As shown in Figure 6(b), a lemon-type vector distribution emerges when the topological charge of the $q$-plate is set to $q$=1/2. Increasing the propagation distance induces a progressive rotation of the global polarization distribution, signifying the dynamic evolution of the Stokes vector fields. This evolution drives periodic transitions between Néel-type and Bloch-type skyrmions, with the transition pathway precisely mapped onto the blue trajectory in Figure 6(a). Figure 6(c) reveals that a star-type vector field generated at $q$=−1/2 facilitates the formation of antiskyrmion textures, whose topological state transitions follow the red trajectory in Figure 6(a).

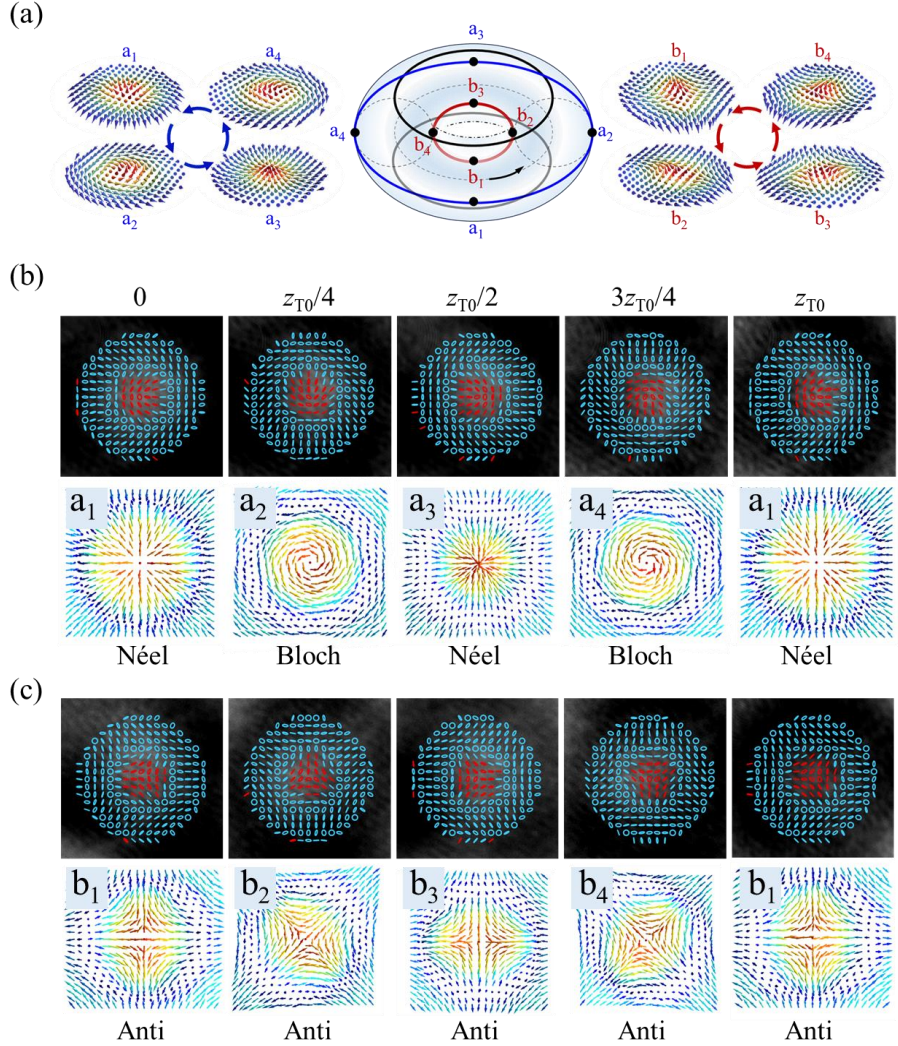

**Figure 6.** (a) skyrmion torus. (b) and (c) show the spatial polarization distribution of the output beam and its corresponding skyrmion texture at different distances. The propagation paths of the skyrmion shapes are indicated by the blue and red lines on the skyrmion torus, respectively.

## 4. Conclusion

In summary, we proposed an LC PBPE based on the checkerboard encoding method, which could realize both amplitude and PB phase modulation. We designed a checkerboard-encoded PBPE for generating a Bessel beam whose polarization state rotates during propagation, named PRB. Correspondingly, we fabricated an LC PBPE and experimentally observed the PRB with a rotation rate of 137.0°/mm within about 13mm. Furthermore, by introducing devices such as a QWP and a *q*-plate, we achieved scalar and vector beams with polarization varying longitudinally along different paths on the Poincaré sphere. In addition, by adjusting the *q*-plate voltage, we construct structured vector beams that can generate optical skyrmions, and realize longitudinally varying skyrmion textures, which further extends the dimensions of modulation of the structured light field. Our results offer a simple method to realize simultaneously the amplitude and phase modulation with a single PBPE and are expected to play an important role in the applications related to polarization control, such as optical processing, microscopic imaging, and particle manipulation.

## Acknowledgments

We gratefully acknowledge financial support from the National Key Research and Development Program of China (2022YFA1404800); National Natural Science Foundation of China (12474338, 12474298, 62305271, 62175200, 12374279, 62475217); Fundamental Research Funds for the Central Universities (5110240015).

## Author Declarations

The authors have no conflicts to disclose.